\documentclass[conference]{IEEEtran}

\usepackage{algorithmic}
\usepackage{graphicx}
\usepackage{textcomp}
\usepackage[dvipsnames]{xcolor}
\usepackage{xspace}
\usepackage{fontawesome}
\usepackage{multirow}
\usepackage{colortbl}
\usepackage{rotating}
\usepackage{tikz}
\usepackage{soul}
\usepackage{makecell}
\usepackage{url}
\usepackage{enumitem}
\usepackage{fancybox}
\usepackage{booktabs}
\usepackage{fontawesome}
\usepackage{setspace}               
\usepackage{pifont}
\usepackage[many]{tcolorbox}    	
\usepackage[caption=false]{subfig}
\usepackage[export]{adjustbox}
\usepackage{amsmath,amsfonts}
\usepackage{MnSymbol}
\usepackage{color,colortbl,booktabs}
\usepackage[noadjust]{cite}
\usepackage{balance}
\usepackage{bm}

\definecolor{main}{HTML}{cccccc}    
\definecolor{sub}{HTML}{000000}     
\newcommand{\darkred}{\color[RGB]{139,0,0}}
\newcommand{\darkgreen}{\color[RGB]{0,100,0}}
\definecolor{darkgreen}{rgb}{0.0, 0.5, 0.0}

\newtcolorbox{boxM}{
	fontupper = \color{black},
	rounded corners,
	arc = 6pt,
	colback = main!80, 
	colframe = main, 
	boxrule = 0pt, 
	bottomrule = 4.5pt,
	enhanced,
	fuzzy shadow = {0pt}{-3pt}{-0.5pt}{0.5pt}{black!35}
}

\newcommand{\mcEval}{\emph{McEval}\xspace}
\newcommand{\mcJava}{\emph{McEval}$_{Java}$\xspace}
\newcommand{\mcPython}{\emph{McEval}$_{Python}$\xspace}
\newcommand{\multipleE}{\emph{MultiPL}-E\xspace}
\newcommand{\multipleJava}{\emph{MultiPL}-E$_{Java}$\xspace}
\newcommand{\multiplePython}{\emph{MultiPL}-$_E{Python}$\xspace}

\newcommand{\ie}{\emph{i.e.,}\xspace}
\newcommand{\eg}{\emph{e.g.,}\xspace}

\newcommand{\etal}{\emph{et~al.}\xspace}
\newcommand{\secref}[1]{Section~\ref{#1}\xspace}

\newcommand{\figref}[1]{Fig.~\ref{#1}\xspace}

\newcommand{\tabref}[1]{Table~\ref{#1}\xspace}

\newcommand*\circled[1]{\tikz[baseline=(char.base)]{
		\node[shape=circle,fill,inner sep=0.8pt] (char) {\textcolor{white}{#1}};}}
\newcommand{\rev}[1]{\textcolor{black}{#1}}
\newboolean{showcomments}
\setboolean{showcomments}{true}

\ifthenelse{\boolean{showcomments}}
{\newcommand{\nb}[2]{
		\fbox{\bfseries\sffamily\scriptsize#1}
		{\sf\small$\blacktriangleright$\textit{#2}$\blacktriangleleft$}
	}
	
}
{\newcommand{\nb}[2]{}
	
}

\AtBeginDocument{%
	\providecommand\BibTeX{{%
			Bib\TeX}}}
\def\BibTeX{{\rm B\kern-.05em{\sc i\kern-.025em b}\kern-.08em
    T\kern-.1667em\lower.7ex\hbox{E}\kern-.125emX}}
\begin{document}

\title{Is Quantization a Deal-breaker?\\Empirical Insights from Large Code Models}

 \author{%
  \IEEEauthorblockN{Saima Afrin}
  \IEEEauthorblockA{%
    \textit{Department of Computer Science}\\
    \textit{William \& Mary}\\
    Williamsburg, VA, USA\\
    safrin@wm.edu
  }
  \and
  \IEEEauthorblockN{Bowen Xu}
  \IEEEauthorblockA{%
    \textit{Department of Computer Science}\\
    \textit{North Carolina State University}\\
    Raleigh, NC, USA\\
    bxu22@ncsu.edu
  }
  \and
  \IEEEauthorblockN{Antonio Mastropaolo}
  \IEEEauthorblockA{%
    \textit{Department of Computer Science}\\
    \textit{William \& Mary}\\
    Williamsburg, VA, USA\\
    amastropaolo@wm.edu
  }
}

\maketitle

\begin{abstract}
The growing scale of large language models (LLMs) not only demands extensive computational resources but also raises environmental concerns due to their increasing carbon footprint.  Model quantization emerges as an effective approach that can reduce the resource demands of LLMs by decreasing parameter precision without substantially affecting performance (\eg 16 bit $\rightarrow$ 4 bit). While recent studies have established quantization as a promising approach for optimizing large code models (LCMs)--a specialized subset of LLMs tailored for automated software engineering--their findings offer only limited insights into its practical implications. Specifically, current investigations only focus on the functional correctness of the code generated by quantized models, neglecting the gap that remains in understanding how quantization impacts critical aspects of quality attributes of the code, such as reliability, maintainability, and security. To bridge this gap, our study investigates the effects of quantization on the qualitative aspects of automatically generated code. We apply Activation-aware Weight Quantization (AWQ) to two widely used code models—CodeLlama and DeepSeekCoder—to generate Java and Python code. Leveraging state-of-the-art static analysis tools, we evaluate software quality metrics and static features, including cyclomatic complexity, cognitive complexity, and lines of code (LoC). Our findings reveal that quantization not only establishes itself as a robust technique capable of withstanding functional challenges-producing code that passes test cases at rates comparable to non-quantized models—but also preserves key qualitative attributes and static features often sought after by developers, such as maintainability and structural complexity. 
\end{abstract}


\section{Introduction}
\label{sec:intro}


In recent years, Software Engineering (SE) has experienced a significant transformation, largely driven by the integration of Artificial Intelligence (AI) techniques into development practices. Among these, Large Language Models (LLMs) have emerged as powerful tools for automating a wide range of SE-related tasks, enabling developers and practitioners to work more efficiently.

Collectively referred to as Large Code Models (LCMs), these models have been successfully applied to various aspects of software development automation, including:
bug fixing, code summarization, code completion, software testing, and others \cite{hou2023large,watson:tosem2022}--showcasing their versatility and impact on modern software engineering.

The rise of LCMs can be largely attributed to advancements in hardware capabilities and the availability of vast training datasets. For example, GitHub \cite{github}, the leading platform for code hosting and software engineering data, reported over 28 million public repositories as of January 2023 \cite{github}. Microsoft has leveraged this extensive dataset to develop GitHub Copilot \cite{copilot}, an AI-powered coding assistant that has already surpassed 1 million users.


However,  the price we pay for such remarkable advancements in automation is the significant computational expense associated with sustainability and environmental impact—factors that are often overlooked by researchers in the field \cite{hou2023large}.
To illustrate the environmental impact of training LCMs, consider StarCoder \cite{li2023starcoder}, a model with 7 billion parameters in its base version. Training this model resulted in 16.68 tons of CO$_{2}$ emissions, which is equivalent to four round-trip flights between Los Angeles and Rome (Based on estimates from \url{https://co2.myclimate.org}). This comparison underscores the significant carbon footprint associated with developing large-scale AI models.

While skeptics may argue that training is a one-time cost—making it seemingly more tolerable—research has shown that the majority of the environmental and computational costs arise after deployment \cite{wu2022sustainable,de2023growing,chien2023reducing}. In particular, model inference—the process of generating responses when prompted—incurs significant energy consumption, often surpassing the cost of initial training over a prolonged period of time. 
In this context, quantization~\cite{liu2023llm, dettmers20218, gholami2022survey} emerges as a powerful optimization technique for compressing LCMs via models' parameters bit-size reduction (\eg 16 bit $\rightarrow$ 4 bit). Additionally, given that LCMs are designed with distinct optimization patterns to enhance software engineering automation, quantization plays a crucial role in boosting efficiency, reducing computational costs, and improving scalability in LCM-driven applications \cite{wei2023towards}.

As LLMs and LCMs continue to grow in size and complexity, so does the interest in making them more efficient and accessible. This has led to a notable surge in publicly available quantized models on platforms like Hugging Face, enabling broader adoption in real-world applications. For instance, the DeepSeek-Coder family \cite{deepseek} is offered in multiple quantized configurations, reflecting the industry's growing emphasis on resource-efficient AI models that strike a balance between performance, scalability, and sustainability.

In the field of code models quantization, pioneering is the work of Wei \etal \cite{wei2023towards}--demonstrating that 8-bit precision quantization can significantly reduce carbon footprint and energy consumption while preserving accuracy (\ie the model’s ability to generate correct code recommendations) and robustness with--virtually--no impact. The results are confirmed in several other investigations--as demonstrated by recent literature \cite{frantar2022gptq, shen2024edgeqat, lin2024awq}.

Although existing research has demonstrated the capability of state-of-the-art quantization methods to handle code-specific tasks, such as source code documentation \cite{afrin2025resource}—a task that requires models to bridge source code with natural language—a significant gap persists in the literature. Specifically, prior studies have largely neglected the quantitative evaluation of quantized models in terms of the quality of the code they produce. 

Code quality, as emphasized in numerous studies \cite{asare2024user, asare2023github, khoury2023secure}, is a cornerstone of effective software development, influencing critical attributes such as maintainability, security, readability, and overall reliability. High-quality code reduces technical debt, boosts developer productivity, and extends the lifespan of software systems. In the context of code generated by LLMs, it is imperative that quantization techniques preserve or enhance code quality rather than degrade it. While a quantized model may generate code that successfully passes test cases, if the resulting code is overly complex, poorly structured, or difficult to maintain, it risks introducing long-term inefficiencies and errors \cite{siddiq2024quality}, thereby diminishing its practical utility. This underscores the need for a comprehensive evaluation framework of quantization's impact on code quality, ensuring that sustainability gains do not come at the expense of software quality and developer effectiveness.


With that in mind, we postulated the following hypothesis as a fundamental assumption of this investigation:\\

\textit{While quantized large code models often maintain the functional correctness of generated code, they may fall short of the high quality standards needed for software development. The inherent information loss from quantization tends to negatively affect maintainability, security, and structural complexity, potentially impacting the overall code quality. This risk may intensify with larger models.}\\

By exploring this hypothesis, we aim to provide a comprehensive understanding of how quantization impacts the quality of code automatically generated by LCMs, extending its evaluation beyond functional correctness to encompass the fundamental dimensions of quality aspects and static properties of the code. Ultimately, our investigation contributes to the current state-of-the-art by offering a comprehensive analysis of quantization and its effect on state-of-the-art LCMs and their ability to produce code not only functionally correct, but code that reflects the quality standard capabilities for LLM-based code generators.

A replication package that includes data, scripts, and documentation is publicly available to promote further research in the field  \cite{replication}.

\section{Background and Related Work}
\label{sec:related}

This section provides the reader with an overview of recent advancements in the development of LLM-based models for code generation and efficiency-based methods that aim to improve the sustainability of such models.

\subsection{Large Code Models for Code Generation}
\label{sec:llm4code}

In recent years, the development and adoption of LLM-based solutions for automating software engineering tasks have made significant progress, that have accelerated the evolution of AI-powered methods for code generation, fundamentally reshaping automation in software engineering. As this AI-driven landscape continues to evolve, researchers persist in developing and refining LLMs for code (\ie LCMs), each with unique strengths in automated code generation.

Several LLM-based models, such as  CodeGen \cite{codegen}, AlphaCode \cite{li2022competition}, PolyCoder \cite{xu2022systematic}, CodeLlama \cite{roziere2023code}, and DeepSeek-Coder \cite{guo2024deepseek}, have demonstrated remarkable capabilities in automated code generation, each excelling in different areas. For example, in the realm of general purpose LLMs--models like Claude \cite{Anthropic_Claude3}, Gemini \cite{gemini}, and GPT-4 \cite{gpt4} have emerged as leading solutions, redefining software engineering while transforming the way developers interact with and integrate automated coding solutions into their workflows. 


For a comprehensive review of LLMs in code generation, we direct readers to the literature survey by Hou \etal \cite{hou2023large}.

\subsection{Improving Sustainability in Large Code Models}
\label{sec:green-ai}
The deployment of billion-parameter LCMs imposes substantial computational demands and considerable energy requirements, often reliant on non-renewable resources, sparking concerns about environmental sustainability \cite{patterson2021carbon, castano2023exploring,strubell2020energy}.

In a recent study, Shi \etal \cite{shi2024efficient} identified four key areas for optimizing and improving the sustainability of LCMs: (i) data reduction, (ii) model-centric, (iii) system-centric, and (iv) program-centric approaches. Each of these strategies offers distinct advantages and is tailored to address specific challenges in enhancing the efficiency of LCMs.

For instance, quantization falls under the model-centric category, as it directly modifies the structure of the network by reducing the precision of its parameters, leading to lower memory usage and faster inference times.

Due to space constraints, we do not delve into all four categories in detail—readers are encouraged to consult Shi \etal \cite{shi2024efficient} for a comprehensive overview. Instead, we focus on model compression techniques—Quantization.

\noindent \textbf{Quantization} is a model compression technique that aims at reducing the model's memory footprint and consequently, speeding-up the inference, by representing its parameters (weights and/or activations) in a lower-precision format, such as 8-bit integers, instead of the standard 16-bit or 32-bit floating points \cite{gholami2022survey}. Quantization can be applied through two possible strategies: Quantization-Aware Training (QAT) and Post-Training Quantization (PTQ).
\textbf{QAT} \cite{esser2019learned} integrates quantization during training, necessitating a complete model retraining. Specific adaptations for LLMs, such as \textit{LLM-QAT} \cite{liu2023llm} and \textit{EdgeQAT} \cite{shen2024edgeqat}, can usually be applied to LLMs having limited size. Indeed, the high GPU time and memory demands render QAT impractical for large-scale LLMs, making PTQ a preferable alternative. \textbf{PTQ} \cite{cai2020zeroq} converts pre-trained models to fixed-point networks without revisiting the initial training stages. This method requires the model parameters to be adjusted to minimize quantization errors using a calibration dataset, but such an ``additional training'' is quite cheap to run.
PTQ techniques can be further divided into weights-only and weight-activation quantization.
\textit{Weights-only} quantization, such as \textit{GPTQ} \cite{frantar2022gptq} and strategies in \textit{PB-LLM} \cite{shang2023pb} and \textit{SpQR} \cite{dettmers2023spqr}, focuses on minimizing precision loss by adjusting weight bit-widths and applying scale transformations for critical weight channels. \textit{Weight-activation} quantization, exemplified by \textit{SmoothQuant} \cite{xiao2023smoothquant}  and \textit{LLM.int8()} \cite{dettmers2022gpt3}, compresses both weights and activations, utilizing techniques like mixed-precision decomposition and channel-wise scaling. Activation aware Weight Quantization (AWQ) \cite{lin2024awq} — a \textit{Weight quantization} technique, takes activation importance into account during the weight quantization process.

\vspace{-1.5pt}

\subsection{Quality of LLMs-based Generated Code}
\label{sec:quality}
A growing body of research has explored the quality and reliability of code produced by Large Language Models (LLMs), particularly in the context of security, maintainability, and correctness. Perry \etal \cite{perry2023users} investigated the security risks associated with LLM-generated code, identifying a higher incidence of vulnerabilities compared to human-written code. Similarly, Sandoval \etal \cite{sandoval2023lost} reported a 10\% increase in security flaws within LLM-generated C code, further emphasizing concerns regarding the robustness of AI-assisted coding.
Mastropaolo \etal \cite{Mastropaolo:icse2023} examined GitHub Copilot's ability to interpret semantically equivalent natural language descriptions of code, finding that it successfully generated code recommendations in only 46\% of cases, with a 28\% correctness gap between different yet equivalent prompts. Asare et al. \cite{asare2024user} observed that while Copilot enhanced security in complex programming tasks, its impact on simpler coding problems was minimal.
Further evaluation by Yetiştiren et al. \cite{yetiştiren2023evaluatingcodequalityaiassisted} assessed the security, maintainability, and reliability of code produced by ChatGPT \cite{gpt4}, GitHub Copilot \cite{copilot}, and Amazon CodeWhisperer \cite{codewhisperer}, employing SonarQube \cite{sonarcloud} as an analysis tool. Additionally, Asare et al. \cite{asare2023github} found that Copilot replicated security vulnerabilities in 33.3\% of cases, while successfully mitigating 25.5\% of them.
Khoury et al. \cite{khoury2023secure} investigated ChatGPT’s security performance across multiple programming languages, demonstrating frequent failures to meet security standards, though iterative prompting led to improved results. 
Although these studies have analyzed LLM-generated code quality from multiple perspectives, including security, maintainability, and reliability, to the best of our knowledge, there is no relevant research that examines these attributes in quantized model-generated code. This gap underscores the need for further investigation into how quantization techniques impact static features and code quality attributes of the automatically generated code.


\section{Study Methodology} 
\label{sec:design}

 The primary objective of this study is to comprehensively examine the impact of the quantization technique on the quality of code generated by LCMs. Despite quantization carrying certain benefits, it is crucial to understand whether those gain comes at the cost of degraded code quality or not. To explore these concerns, we formulate the research question underpinning our investigation (RQ):

\begin{itemize}
	\item \textbf{RQ$_1$:} \emph{How does quantization impact the quality of automatically generated code? } 
    
    In RQ$_1$, we evaluate the quality attributes of automatically generated code, including maintainability, readability, complexity, and the presence of code smells, using state-of-the-art static analysis tools. 
    The objective is to address a fundamental concern rooted in the assumption that the loss of information — an inherent byproduct of the quantization process-may impact code quality in ways that are not yet fully understood. 

    Building on this premise, we further examine how model size influences the severity of this quality degradation. Since quantization compresses models significantly, larger models are hypothesized to suffer greater precision loss than their smaller counterparts. To explore this, we analyze different quantized variants within the same model family, comparing the quality of their generated code to understand the relationship between model size and post-quantization quality.    \\

	  
    
\end{itemize}

\smallskip

Moving forward, we analyze the behavior of two prominent code model families-CodeLlama \cite{roziere2023code} and DeepSeek-Coder \cite{deepseek}-using two widely recognized benchmarks and four static analysis tools. These static analysis tools are developed to evaluate code quality across multiple programming languages. In this regard, we specifically focus on Java and Python--being consistent with relevant research in the field \cite{codellama2, ren2024reflectioncoder} while enhancing the generalizability of our findings.


\subsection{Code Generation Models}
\label{sub:models}

In this work, we selected two state-of-the-art code model families: CodeLlama \cite{roziere2023code} and DeepSeek-Coder \cite{deepseekcoder}. These models have been widely studied for various code-related tasks, including code generation, and have consistently demonstrated strong performance in prior research \cite{li2023structured, coignion2024performance, ren2024reflectioncoder}. Both families offer multiple variants, including instruction-tuned and non-instruction-tuned versions. We utilize the instruction-tuned variant to align with relevant research  \cite{afrin2025resource, yuan2023evaluating, li2023instructcoder}. 


\smallskip

\subsubsection{CodeLlama \cite{roziere2023code}} is an open-source LLM family designed for coding tasks. Built upon the foundational Llama-2 model \cite{touvron2023llama}, it has been further trained on a corpus comprising 500 billion tokens, encompassing both natural language and source code. CodeLlama is available in several specialized variants \footnote{\url{https://huggingface.co/codellama}}, each tailored for distinct applications such as general-purpose coding model and instruct variant, optimized for instruction-following tasks.
The model family includes multiple parameter configurations, ranging from 7B to 70B, with all versions publicly accessible. 


\smallskip

\subsubsection{DeepSeek-Coder \cite{deepseekcoder}} represents another leading family of open-source LLMs, with model sizes spanning $\sim$1B to 33B parameters. 
Each model is available in two distinct operational modes: instruct, optimized for instruction tuning, and base, which serves as a foundational, general-purpose model. 
These models have undergone extensive pre-training on 2 trillion tokens, including a substantial volume of code-related data, enabling them to achieve state-of-the-art performance in automating software engineering tasks. Remarkably, DeepSeek-Coder has outperformed significantly larger models, including GPT-3.5 \cite{brown2020language}. Moreover, its mid-sized variant (6.7B parameters) has exhibited competitive performance relative to the 33B parameter version of CodeLlama.

\subsection{Selected Quantization Method }
\label{sub:technique}
\textbf{Activation-Aware Weight Quantization (AWQ)} \cite{lin2024awq} is a PTQ method that prioritizes model activations rather than focusing solely on weights. Unlike traditional quantization techniques, AWQ identifies and preserves salient weights—the most critical parameters that significantly impact LLM performance—by analyzing activations rather than the weights themselves.
This approach is based on the key observation that not all weights contribute equally to model accuracy, and safeguarding only 1\% of the most influential weights can substantially reduce quantization error. To achieve this, AWQ leverages activation magnitude to identify this subset of critical weights and employs per-channel scaling to minimize the quantization error introduced during compression. We selected AWQ for our study based on several key considerations. First, AWQ enables quantization down to 4-bit or lower while maintaining competitive accuracy, allowing us to examine the impact of low-precision model quantization on LLM-based generated code. Additionally, when evaluated on the MBPP code generation dataset, AWQ significantly outperformed earlier methods such as GPTQ and RTN (round-to-nearest) quantization in terms of accuracy retention \cite{lin2024awq}. Ultimately, AWQ has also been demonstrated to achieve superior inference speed compared to GPTQ and AutoGPTQ, further enhancing its efficiency for large-scale deployment.

\subsection{Evaluation Dataset}
\label{sub:dataset}
To generate code utilizing the selected code models, we selected two state-of-the-art benchmark datasets designed to evaluate the model's code generation capability across two programming languages: Python and Java. These benchmarks provide a diverse and challenging set of tasks, enabling a comprehensive analysis of model effectiveness on generating code across different levels of complexity and programming paradigms \cite{humaneval,wei2023towards,cassano2023knowledge}.

\subsubsection{\multipleE \cite{cassano:tse2023}}

\multipleE is a comprehensive code generation benchmark introduced by Cassano \etal \cite{cassano:tse2023}, incorporating two widely used datasets: HumanEval \cite{humaneval} and MBPP \cite{mbpp}, along with their translations into 18 programming languages. For this study, we specifically adopt the HumanEval dataset, as it has been identified as more challenging than MBPP \cite{mbpp}.
Originally, HumanEval consisted of 164 hand-crafted Python prompts, each accompanied by a canonical solution and a set of unit test cases for evaluating model-generated outputs. Within the \multipleE benchmark, 161 of these Python tasks have been retained, of which 158 have been successfully translated into Java. Notably, three tasks were excluded from translation due to their reliance on Python-specific syntax, which lacks a direct equivalent in Java.

\subsubsection{\mcEval \cite{chai2024mceval}}
\mcEval is a diverse and human-annotated collection of coding tasks spanning multiple programming languages. Each task is structured with a function signature, a detailed problem specification provided as a docstring, a set of test cases, and a difficulty classification categorized into easy, medium, and hard levels.
For this study, we extract the Python and Java subsets of \mcEval, initially comprising 50 Python and 53 Java tasks. To ensure dataset integrity and eliminate redundancy, we exclude eight Python tasks that overlap with HumanEval, reducing the final set to 42 Python samples. The resulting dataset comprises 23 easy, 10 medium, and 9 hard problems in Python, while the Java subset includes 30 easy, 13 medium, and 10 hard tasks.

\subsection{Static Analysis Tools \& Code Quality}
\label{sub:tools}
The quality of the generated code can be measured by accounting for various dimensions--including syntax validity, coding style, code length, code smells, reliability, maintainability, and security \cite{siddiq2023generate, yetiştiren2023evaluatingcodequalityaiassisted}. Various static analysis tools provide assessments focusing on different quality attributes. In this study, we employ four widely recognized static analysis tools for code quality evaluation: SonarCloud \cite{sonarcloud}, which supports analysis for both Python and Java, Pylint \cite{pylint} and Flake8 \cite{checkstyle} for Python code assessment, and PMD \cite{pmd} for Java code evaluation.
The selection of these static analysis tools is motivated by their ability to provide diverse perspectives on code quality across the two languages, thereby improving the reliability and generalizability of our evaluation through a more comprehensive assessment of the generated code \cite{siddiq2024quality, liu2024refining, kharma2025security, liu2023your}.

It is worth noting that we excluded CheckStyle \cite{checkstyle} for Java, as it heavily depends on the specific style configurations set by development teams, which we believed could introduce unwanted bias into the interpretation of the results.

\begin{itemize}
\item \textbf{SonarCloud \cite{sonarcloud}} is a cloud-based static analysis tool designed to assist developers in identifying and resolving code quality issues across various programming languages. 

\item \textbf{Pylint \cite{pylint}} is a widely used static analysis tool for enforcing coding standards and detecting issues in Python code \cite{siddiq2024quality, liu2024refining}. 

\item \textbf{Flake8 \cite{flake8}}  is a popular static analysis tool for Python that detects syntax errors, enforces style conventions, and analyzes code complexity. 

\item \textbf{PMD \cite{pmd}} is a static analysis tool for Java that detects best practice violations, code complexity, performance inefficiencies, and maintainability issues. 


\end{itemize}

\section{Implementation and Evaluation}
\label{sec:implementation}
\subsection{Evaluation Pipeline}
Our evaluation framework for assessing the quality of code generated by LCMs consists of two primary stages. First, we employ selected benchmarks to generate code using the baseline (\ie full-precision model) and its quantized counterparts. The functional correctness of the generated code is systematically evaluated by computing the pass@k \cite{chen:arxiv2021} metric, which measures the probability that at least one of the top-k generated code completions successfully passes all test cases. This metric is widely recognized for assessing the reliability of AI-assisted code generation and is currently regarded as the ``de facto standard'' for evaluating LCMs. To ensure a controlled execution environment, the generated code is run within a secure Docker environment, where test cases are leveraged to assess correctness. We calculate the pass@1 metric, which evaluates the model's capability to generate a correct solution on its first attempt. This is done following the settings adopted in established research in the field \cite{deepseekcoder,codellama,wizardcoder}.
Finally, we set a temperature of 0.8 for CodeLlama, following the recommendations of Roziere \etal \cite{roziere2023code}. For DeepSeek-Coder, we adhered to the official configuration guidelines and set the temperature to 0.7. In line with related studies \cite{wang2023review, fakhoury2024llm}, we limit the number of input tokens to 1024 for both models.

\subsection{Experimental Environment}

All experiments were conducted on a server running Ubuntu 22.04.5 LTS (GNU/Linux 5.15.0-125-generic x86\_64), equipped with two Nvidia L40S GPUs, each with 48GB of dedicated graphics memory.

\subsection{Model Quantization}

As detailed in Section \secref{sub:technique}, we applied Activation-Aware Weight Quantization (AWQ) to enable 4-bit model quantization. To maintain consistency and accuracy, we utilized pre-quantized AWQ models available on Hugging Face\footnote{\url{https://huggingface.co/TheBloke}}, ensuring adherence to the platform’s recommended quantization guidelines\footnote{\url{https://huggingface.co/docs/transformers/v4.35.0/main_classes/quantization}}.

To maintain uniformity across all models, we used Hugging Face’s implementations for both the base and quantized versions, ensuring a standardized evaluation framework.


\subsection{Employing Code Analysis Tools }

The evaluation of the qualitative properties of the automatically generated code begins with the application of the selected static analysis tools that are: Pylint, Flake8, PMD, and SonarCloud.
For Python, we use Pylint and Flake8, while for Java, only PMD due to the exclusion of Checkstyle as discussed in \secref{sub:tools}.

These tools perform static analysis by examining each line of code, identifying quality-related issues based on predefined criteria across multiple dimensions. Additionally, we follow established practices \cite{kharma2025security} in the field by integrating SonarCloud for automated quality assessment across multiple languages.  Its ability to provide a unified environment for evaluating code quality makes it an ideal candidate for our study. 

\subsection{Quality Metrics Attributes }

\noindent\textbf{Quality Metrics of SonarCloud:}
The evaluation focuses on five key software quality indicators:

\begin{itemize}
    \item \textbf{Reliability:} Assesses the robustness of the code under predefined conditions. To quantify reliability, we analyzed bug density in the code generated by both full-precision and quantized models.
    
    \item \textbf{Maintainability:} Measures the ease of understanding, modifying, and extending the code over time. 
    
    \item \textbf{Lines of Code (LoC):} Represents the total number of non-whitespace lines in a program. LoC serves as a predictive metric for assessing both effort and maintainability. 
    
    \item \textbf{Cyclomatic Complexity (CyC):} Evaluates code complexity by constructing a control flow graph (CFG), where $E$ represents edges, $N$ denotes nodes, and $Q$ accounts for connected components. The Cyclomatic Complexity $(M)$ is calculated as: $M = E + 2Q - N$

    \item \textbf{Cognitive Complexity (CoC):} Measures the difficulty of understanding the code. Unlike traditional complexity metrics, CoC considers structural elements such as control flow and nesting, and is computed as:
    \[
    C = C_{\text{base}} + \sum_{i=1}^{n} n_c
    \]
    where $n_c$ represents increases in complexity due to nesting and conditionals \cite{munoz2020empirical}.
\end{itemize}


\noindent\textbf{Quality Metrics of Pylint:} 
Pylint’s rule set is categorized into five distinct levels of increasing complexity, each represented by a specific letter:
\begin{itemize}
    \item \textbf{C (Convention)}: Identifies deviations from PEP 8 guidelines, including issues related to indentation, naming conventions, and whitespace, ensuring code consistency and readability.
    \item \textbf{R (Refactor):} Detects suboptimal coding practices such as redundant code, deep nesting, and overly complex functions, providing recommendations to enhance maintainability.
    \item \textbf{W (Warning):} Highlights potential concerns, including unused imports, unused variables, or shadowing of built-in functions, which may introduce unintended behavior.
    \item \textbf{E (Error):} Reports probable bugs, such as syntax errors, invalid arguments, and undefined variables, which could lead to runtime failures.
\end{itemize}

In our study, we focused on the Convention (C), Refactor (R), Warning (W), and Error (E) categories, as they provide essential insights into code style adherence, maintainability, potential issues, and critical errors in the generated code. For the sake of presenting the results, we discarded Fatal as it has never been across the two benchmarks.\\

\noindent \textbf{Quality Metrics of Flake8:} We choose Flake8 for its ability to analyze code style, detect syntax errors, and promote simplicity, ultimately enhancing readability and maintainability. 
Flake8 categorizes issues into distinct groups:

\begin{itemize}
    \item \textbf{E (Error)}: Flags syntax errors and violations of Python’s syntax rules, such as missing colons, incorrect indentation, or misplaced parentheses.
    \item \textbf{W (Warning):} Highlights potential issues in code formatting and structure that may not result in immediate errors but could affect maintainability.
    \item \textbf{F (PyFlakes Errors):} Detects logic errors, including undefined variables, unused imports, and redefined functions or variables.
\end{itemize}

As we did for Pylint, we avoid reporting the metrics for which no change, either positive or negative, has been observed. In other words, if the outputs of these metrics are 0 across benchmarks, we avoid reporting them.  In this case, we exclude the Naming Conventions and the McCabe index. 

\vspace{7pt}
\noindent\textbf{Quality Metrics of PMD:} We leverage PMD’s capabilities to analyze Java code, detecting and categorizing issues based on a predefined ruleset. These categories offer a structured framework for identifying and improving various aspects of code quality, maintainability, and efficiency:

\begin{itemize}
    \item \textbf{Best Practices:} This category helps identify code that may violate fundamental design principles or introduce inefficiencies that could make maintenance difficult.
    \item \textbf{Code Style:} Ensures adherence to specific coding conventions, focusing on brace formatting and naming rules to maintain code consistency.
    \item \textbf{Design:} Identifies design issues, such as deeply nested if statements that will affect maintainability.
    \item \textbf{Error Prone:} Detects constructs that are broken, highly confusing, or likely to cause runtime errors, emphasizing the importance of code readability to prevent such issues.
    \item \textbf{Multi-threading:} Flags potential issues when handling multiple threads of execution to ensure proper synchronization and avoid concurrency-related problems.
    \item \textbf{Performance:} Identifies suboptimal code that may degrade performance, highlighting inefficiencies that could lead to slower execution times and increased resource consumption.
\end{itemize}

\subsection{Analysis Methodology}
To systematically address our research questions (RQs), we task the LCMs underpinning our investigation with a total of $\sim$200 queries (\ie code generation problems). Each model is queried under two distinct conditions: (i) using the full-precision (FP) version, and (ii) using the 4-bit AWQ-based quantized counterpart. This results in an equal split of queries between the two model configurations, ensuring a balanced comparison. For each query, we capture and store the generated code output (\eg from DeepSeek-Coder 1.3B FP) and analyze it using the collection of static analysis tools detailed above. The same procedure is repeated for its quantized equivalent (\ie  DeepSeek-Coder 1.3B AWQ-4bit) to assess the potential differences introduced by quantization.

Next, we focus on key quality metrics extracted from each static analysis tool and conduct a Wilcoxon signed-rank test \cite{wilcoxon} to compare the distributions of various metrics obtained by contrasting quantized vs. full-precision models. \rev{The statistical analysis is conducted at the granularity of individual tasks, where each task serves as a paired observation comprising the number of static analysis issues identified in both the full-precision and quantized model outputs.} To account for multiple comparisons, we apply Holm's correction procedure \cite{Holm1979a} that adjusts the $p$-values to ensure statistical robustness. Additionally, we compute the paired Cliff’s delta effect size \cite{Cliff:2005} to quantify the magnitude of the observed differences.


\subsection{Manually assessing the quality of the identifiers, string literals, and code understandability}
\label{sub:manual-analysis}
To further investigate the impact of quantization on code generation models, we conduct a manual assessment of the identifiers and literals used in the generated code. This evaluation spans across the two benchmarks and programming languages,  providing deeper insights into how quantization affects consistency and readability.

To assess these aspects, we perform a manual analysis of identifier and literal choices in code generated by quantized (AWQ-4bit) LCMs. Our evaluation is based on a randomly selected set of 50 predictions, exclusively sampled from the largest model of the two families, being CodeLlama 34B.

We compute the total number of issues as the sum of SonarCloud metrics, including--\emph{Security, Reliability, Maintainability, Cyclomatic Complexity, and Cognitive Complexity}. In total, we analyze 100 predictions, with 50 samples per experimental configuration (\ie for each programming language within a fixed benchmark).

Each sampled prediction is inspected by two authors of the paper, who independently classify the predictions. To measure consistency between their assessments, we compute the inter-rater agreement using Cohen’s Weighted $\kappa$, ensuring the reliability of our manual evaluation \cite{cohen1960coefficient}--that builds on a 3-point Likert Scale \cite{joshi2015likert} evaluation of two fundamental dimensions of source code.

\begin{itemize}

    \item \textbf{Consistency}: Check if identifiers and literals adhere to standard naming conventions and maintain consistency across various code segments.
    
	\item \textbf{Readability:} The extent to which quantization affects the understandability of generated code.

\end{itemize}


Due to space constraints, our replication package includes a comprehensive table detailing the full evaluation criteria, with assessments classified as follows: 1 (Poor), 2 (Acceptable), and 3 (Good).
\rev{The results indicate that the majority of code generated by quantized models (82\%) were rated as either ``Acceptable'' or ``Good'', suggesting that quantization typically preserves key aspects of code quality. Specifically, 27\% of the samples were rated ``Good'' (minimal-to-no issues; clear, concise, and well-structured code), 55\% were rated ``Acceptable'' (minor concerns not obstructing comprehension or maintainability), and 18\% were rated ``Poor'' (significant consistency or readability issues that may require revision).}



\section{Results and Discussion}
\label{sec:results}

\begin{table*}[ht!]
     \centering
     \caption{
Quality Metrics of Different Models Benchmarked on \emph{MultipLE}-Python (\ie HumanEval) and \emph{McEval}-Python. Cyc refers to Cyclomatic Complexity, while CoC denotes Cognitive Complexity}
     \normalsize
     \label{tab:model-performance-py}
     \resizebox{\linewidth}{!}{
         \begin{tabular}{llcc|cccccc|cccc|ccc}
             \toprule
             \textbf{Dataset} & \textbf{Model} & \textbf{Precision} & \textbf{Pass@1} & \multicolumn{6}{c|}{\textbf{SonarCloud Metrics}} & \multicolumn{4}{c|}{\textbf{Pylint}} & \multicolumn{3}{c}{\textbf{Flake8}} \\
             \cmidrule(lr){5-10} \cmidrule(lr){11-14} \cmidrule(lr){15-17}
             & & & & \textbf{LoC} & \textbf{Security} & \textbf{Reliability} & \textbf{Maintainability} & \textbf{CyC} & \textbf{CoC} & \textbf{Error} & \textbf{Warning} & \textbf{Convention} & \textbf{Refactor} & \textbf{Error} & \textbf{Warning} & \textbf{PyFlakes} \\
             \midrule
             
             \multirow{12}{*}{\centering \emph{MultipLE}-Python} & \multirow{2}{*}{CodeLllama 7B} & \cellcolor[gray]{.75} 16 bit & \cellcolor[gray]{.75} 0.30 & \cellcolor[gray]{.75} 1078 & \cellcolor[gray]{.75} 0 & \cellcolor[gray]{.75} 5 & \cellcolor[gray]{.75} 47 & \cellcolor[gray]{.75} 491 & \cellcolor[gray]{.75} 522 & \cellcolor[gray]{.75} 36 & \cellcolor[gray]{.75} 28 & \cellcolor[gray]{.75} 393 & \cellcolor[gray]{.75} 32 & \cellcolor[gray]{.75} 304 & \cellcolor[gray]{.75} 130 & \cellcolor[gray]{.75} 32 \\
             
             & & \cellcolor[gray]{.95} 4 bit & \cellcolor[gray]{.95} 0.27 & \cellcolor[gray]{.95} 1118 $\darkgreen\uparrow$ & \cellcolor{red!25} 1 & \cellcolor{green!20} 2 & \cellcolor{green!20} 28 & \cellcolor{green!25} 470 & \cellcolor{red!25} 537 & \cellcolor{green!20} 17 & \cellcolor{green!25} 27 & \cellcolor{green!25} 390 & \cellcolor{yellow!25} 32 & \cellcolor{red!25} 309 & \cellcolor{red!25} 136 & \cellcolor{green!20} 10 \\
             
             & \multirow{2}{*}{CodeLllama 13B} & \cellcolor[gray]{.75} 16 bit & \cellcolor[gray]{.75} 0.35 & \cellcolor[gray]{.75} 1130 & \cellcolor[gray]{.75} 0 & \cellcolor[gray]{.75} 1 & \cellcolor[gray]{.75} 29 & \cellcolor[gray]{.75} 451 & \cellcolor[gray]{.75} 409 & \cellcolor[gray]{.75} 7 & \cellcolor[gray]{.75} 56 & \cellcolor[gray]{.75} 395 & \cellcolor[gray]{.75} 25 & \cellcolor[gray]{.75} 305 & \cellcolor[gray]{.75} 129 & \cellcolor[gray]{.75} 8 \\
             
             & & \cellcolor[gray]{.95} 4 bit & \cellcolor[gray]{.95} 0.36 & \cellcolor[gray]{.95} 1180 $\darkgreen\uparrow$ & \cellcolor{red!25} 1 & \cellcolor{red!25} 2 & \cellcolor{red!25} 32 & \cellcolor{red!25} 486 & \cellcolor{red!20} 524 & \cellcolor{red!25} 10 & \cellcolor{green!25} 39 & \cellcolor{red!25} 405 & \cellcolor{red!25} 36 & \cellcolor{red!25} 313 & \cellcolor{red!25} 134 & \cellcolor{red!25} 9 \\
             
             & \multirow{2}{*}{CodeLllama 34B} & \cellcolor[gray]{.75} 16 bit & \cellcolor[gray]{.75} 0.41 & \cellcolor[gray]{.75} 1134 & \cellcolor[gray]{.75} 2 & \cellcolor[gray]{.75} 1 & \cellcolor[gray]{.75} 25 & \cellcolor[gray]{.75} 461 & \cellcolor[gray]{.75} 504 & \cellcolor[gray]{.75} 13 & \cellcolor[gray]{.75} 34 & \cellcolor[gray]{.75} 394 & \cellcolor[gray]{.75} 31 & \cellcolor[gray]{.75} 321 & \cellcolor[gray]{.75} 127 & \cellcolor[gray]{.75} 13 \\
             
             & & \cellcolor[gray]{.95} 4 bit & \cellcolor[gray]{.95} 0.42 & \cellcolor[gray]{.95} 1164 $\darkgreen\uparrow$ & \cellcolor{yellow!25} 2 & \cellcolor{green!25} 0 & \cellcolor{red!25} 26 & \cellcolor{red!25} 492 & \cellcolor{red!25} 525 & \cellcolor{green!20} 6 & \cellcolor{green!20} 15 & \cellcolor{red!25} 398 & \cellcolor{green!25} 23 & \cellcolor{red!25} 330 & \cellcolor{green!25} 124 & \cellcolor{green!25} 8 \\
             
             \cline{2-17}
             
             & \multirow{2}{*}{DeepSeek-Coder-1.3B} & \cellcolor[gray]{.75} 16 bit & \cellcolor[gray]{.75} 0.56 & \cellcolor[gray]{.75} 1242 & \cellcolor[gray]{.75} 1 & \cellcolor[gray]{.75} 1 & \cellcolor[gray]{.75} 19 & \cellcolor[gray]{.75} 527 & \cellcolor[gray]{.75} 540 & \cellcolor[gray]{.75} 0 & \cellcolor[gray]{.75} 20 & \cellcolor[gray]{.75} 414 & \cellcolor[gray]{.75} 28 & \cellcolor[gray]{.75} 330 & \cellcolor[gray]{.75} 124 & \cellcolor[gray]{.75} 1 \\
             
             & & \cellcolor[gray]{.95} 4 bit & \cellcolor[gray]{.95} 0.52 & \cellcolor[gray]{.95} 1180 $\darkred\downarrow$ & \cellcolor{yellow!25} 1 & \cellcolor{yellow!25} 1 & \cellcolor{green!25} 17 & \cellcolor{green!25} 494 & \cellcolor{red!25} 543 & \cellcolor{red!25} 3 & \cellcolor{red!25} 29 & \cellcolor{green!25} 408 & \cellcolor{yellow!25} 28 & \cellcolor{green!25} 323 & \cellcolor{red!25} 136 & \cellcolor{red!25} 4 \\
             
             & \multirow{2}{*}{DeepSeek-Coder-6.7B} & \cellcolor[gray]{.75} 16 bit & \cellcolor[gray]{.75} 0.69 & \cellcolor[gray]{.75} 1240 & \cellcolor[gray]{.75} 1 & \cellcolor[gray]{.75} 3 & \cellcolor[gray]{.75} 18 & \cellcolor[gray]{.75} 519 & \cellcolor[gray]{.75} 576 & \cellcolor[gray]{.75} 1 & \cellcolor[gray]{.75} 9 & \cellcolor[gray]{.75} 397 & \cellcolor[gray]{.75} 32 & \cellcolor[gray]{.75} 326 & \cellcolor[gray]{.75} 128 & \cellcolor[gray]{.75} 2 \\
             
             & & \cellcolor[gray]{.95} 4 bit & \cellcolor[gray]{.95} 0.63 & \cellcolor[gray]{.95} 1237 $\darkred\downarrow$ & \cellcolor{yellow!25} 1 & \cellcolor{red!25} 4 & \cellcolor{red!25} 19 & \cellcolor{green!25} 511 & \cellcolor{green!25} 520 & \cellcolor{green!25} 0 & \cellcolor{red!20} 30 & \cellcolor{red!25} 403 & \cellcolor{green!25} 31 & \cellcolor{red!25} 340 & \cellcolor{red!25} 129 & \cellcolor{yellow!25} 2 \\
             
             & \multirow{2}{*}{DeepSeek-Coder-33B} & \cellcolor[gray]{.75} 16 bit & \cellcolor[gray]{.75} 0.64 & \cellcolor[gray]{.75} 1110 & \cellcolor[gray]{.75} 1 & \cellcolor[gray]{.75} 2 & \cellcolor[gray]{.75} 17 & \cellcolor[gray]{.75} 455 & \cellcolor[gray]{.75} 453 & \cellcolor[gray]{.75} 5 & \cellcolor[gray]{.75} 51 & \cellcolor[gray]{.75} 392 & \cellcolor[gray]{.75} 25 & \cellcolor[gray]{.75} 327 & \cellcolor[gray]{.75} 124 & \cellcolor[gray]{.75} 5 \\
             
             & & \cellcolor[gray]{.95} 4 bit & \cellcolor[gray]{.95} 0.58 & \cellcolor[gray]{.95} 1138 $\darkgreen\uparrow$ & \cellcolor{green!25} 0 & \cellcolor{red!25} 1 & \cellcolor{red!25} 29 & \cellcolor{red!25} 466 & \cellcolor{red!25} 491 & \cellcolor{yellow!25} 5 & \cellcolor{green!25} 43 & \cellcolor{green!25} 386 & \cellcolor{green!25} 24 & \cellcolor{green!25} 312 & \cellcolor{green!25} 122 & \cellcolor{red!25} 7 \\
             \midrule
            
             \multirow{12}{*}{\centering \emph{McEval}-Python} & \multirow{2}{*}{CodeLllama 7B} & \cellcolor[gray]{.75} 16 bit & \cellcolor[gray]{.75} 0.11 & \cellcolor[gray]{.75} 1009 & \cellcolor[gray]{.75} 0 & \cellcolor[gray]{.75} 1 & \cellcolor[gray]{.75} 35 & \cellcolor[gray]{.75} 232 & \cellcolor[gray]{.75} 251 & \cellcolor[gray]{.75} 5 & \cellcolor[gray]{.75} 26 & \cellcolor[gray]{.75} 347 & \cellcolor[gray]{.75} 9 & \cellcolor[gray]{.75} 455 & \cellcolor[gray]{.75} 125 & \cellcolor[gray]{.75} 11 \\
             
             & & \cellcolor[gray]{.95} 4 bit & \cellcolor[gray]{.95} 0.11 & \cellcolor[gray]{.95} 1066 $\darkgreen\uparrow$ & \cellcolor{yellow!25} 0 & \cellcolor{yellow!25} 1 & \cellcolor{red!25} 38 & \cellcolor{green!25} 229 & \cellcolor{green!25} 210 & \cellcolor{green!25} 5 & \cellcolor{green!25} 20 & \cellcolor{red!20} 412 & \cellcolor{green!25} 6 & \cellcolor{red!25} 497 & \cellcolor{red!25} 139 & \cellcolor{red!25} 13 \\
             
             & \multirow{2}{*}{CodeLllama 13B} & \cellcolor[gray]{.75} 16 bit & \cellcolor[gray]{.75} 0.21 & \cellcolor[gray]{.75} 1105 & \cellcolor[gray]{.75} 0 & \cellcolor[gray]{.75} 0 & \cellcolor[gray]{.75} 26 & \cellcolor[gray]{.75} 248 & \cellcolor[gray]{.75} 274 & \cellcolor[gray]{.75} 9 & \cellcolor[gray]{.75} 17 & \cellcolor[gray]{.75} 386 & \cellcolor[gray]{.75} 8 & \cellcolor[gray]{.75} 519 & \cellcolor[gray]{.75} 138 & \cellcolor[gray]{.75} 11 \\
             
             & & \cellcolor[gray]{.95} 4 bit & \cellcolor[gray]{.95} 0.22 & \cellcolor[gray]{.95} 1040 $\darkred\downarrow$ & \cellcolor{yellow!25} 0 & \cellcolor{yellow!25} 0 & \cellcolor{green!25} 23 & \cellcolor{green!25} 225 & \cellcolor{green!25} 214 & \cellcolor{green!25} 6 & \cellcolor{green!25} 13 & \cellcolor{red!25} 400 & \cellcolor{red!25} 9 & \cellcolor{green!25} 504 & \cellcolor{green!25} 128 & \cellcolor{green!20} 6 \\
             
             & \multirow{2}{*}{CodeLllama 34B} & \cellcolor[gray]{.75} 16 bit & \cellcolor[gray]{.75} 0.21 & \cellcolor[gray]{.75} 1088 & \cellcolor[gray]{.75} 0 & \cellcolor[gray]{.75} 1 & \cellcolor[gray]{.75} 30 & \cellcolor[gray]{.75} 246 & \cellcolor[gray]{.75} 250 & \cellcolor[gray]{.75} 2 & \cellcolor[gray]{.75} 13 & \cellcolor[gray]{.75} 371 & \cellcolor[gray]{.75} 10 & \cellcolor[gray]{.75} 506 & \cellcolor[gray]{.75} 137 & \cellcolor[gray]{.75} 9 \\
             
             & & \cellcolor[gray]{.95} 4 bit & \cellcolor[gray]{.95} 0.26 & \cellcolor[gray]{.95} 1114 $\darkgreen\uparrow$ & \cellcolor{yellow!25} 0 & \cellcolor{yellow!25} 0 & \cellcolor{red!25} 32 & \cellcolor{red!25} 254 & \cellcolor{red!25} 269 & \cellcolor{green!25} 1 & \cellcolor{green!25} 12 & \cellcolor{red!25} 404 & \cellcolor{green!25} 8 & \cellcolor{green!25} 503 & \cellcolor{green!25} 131 & \cellcolor{green!25} 6 \\
             
             \cline{2-17}
             
             & \multirow{2}{*}{DeepSeek-Coder-1.3B} & \cellcolor[gray]{.75} 16 bit & \cellcolor[gray]{.75} 0.19 & \cellcolor[gray]{.75} 1030 & \cellcolor[gray]{.75} 0 & \cellcolor[gray]{.75} 0 & \cellcolor[gray]{.75} 23 & \cellcolor[gray]{.75} 220 & \cellcolor[gray]{.75} 206 & \cellcolor[gray]{.75} 7 & \cellcolor[gray]{.75} 17 & \cellcolor[gray]{.75} 362 & \cellcolor[gray]{.75} 11 & \cellcolor[gray]{.75} 484 & \cellcolor[gray]{.75} 134 & \cellcolor[gray]{.75} 12 \\
             
             & & \cellcolor[gray]{.95} 4 bit & \cellcolor[gray]{.95} 0.16 & \cellcolor[gray]{.95} 884 $\darkred\downarrow$ & \cellcolor{yellow!25} 0 & \cellcolor{red!25} 1 & \cellcolor{green!20} 12 & \cellcolor{green!25} 185 & \cellcolor{green!20} 149 & \cellcolor{red!20} 12 & \cellcolor{green!20} 10 & \cellcolor{red!25} 365 & \cellcolor{green!20} 4 & \cellcolor{green!25} 432 & \cellcolor{green!25} 132 & \cellcolor{green!25} 10 \\
             
             & \multirow{2}{*}{DeepSeek-Coder-6.7B} & \cellcolor[gray]{.75} 16 bit & \cellcolor[gray]{.75} 0.40 & \cellcolor[gray]{.75} 1100 & \cellcolor[gray]{.75} 0 & \cellcolor[gray]{.75} 0 & \cellcolor[gray]{.75} 26 & \cellcolor[gray]{.75} 247 & \cellcolor[gray]{.75} 270 & \cellcolor[gray]{.75} 1 & \cellcolor[gray]{.75} 10 & \cellcolor[gray]{.75} 370 & \cellcolor[gray]{.75} 9 & \cellcolor[gray]{.75} 498 & \cellcolor[gray]{.75} 135 & \cellcolor[gray]{.75} 5 \\
             
             & & \cellcolor[gray]{.95} 4 bit & \cellcolor[gray]{.95} 0.38 & \cellcolor[gray]{.95} 1071 $\darkred\downarrow$ & \cellcolor{yellow!25} 0 & \cellcolor{yellow!25} 0 & \cellcolor{red!25} 28 & \cellcolor{red!25} 252 & \cellcolor{green!25} 268 & \cellcolor{yellow!25} 1 & \cellcolor{green!25} 12 & \cellcolor{red!25} 418 & \cellcolor{green!25} 6 & \cellcolor{red!25} 509 & \cellcolor{red!25} 137 & \cellcolor{yellow!25} 5 \\
             
             & \multirow{2}{*}{DeepSeek-Coder-33B} & \cellcolor[gray]{.75} 16 bit & \cellcolor[gray]{.75} 0.50 & \cellcolor[gray]{.75} 1054 & \cellcolor[gray]{.75} 0 & \cellcolor[gray]{.75} 0 & \cellcolor[gray]{.75} 23 & \cellcolor[gray]{.75} 227 & \cellcolor[gray]{.75} 248 & \cellcolor[gray]{.75} 0 & \cellcolor[gray]{.75} 5 & \cellcolor[gray]{.75} 417 & \cellcolor[gray]{.75} 7 & \cellcolor[gray]{.75} 490 & \cellcolor[gray]{.75} 138 & \cellcolor[gray]{.75} 4 \\
             
             & & \cellcolor[gray]{.95} 4 bit & \cellcolor[gray]{.95} 0.52 & \cellcolor[gray]{.95} 1025 $\darkred\downarrow$ & \cellcolor{yellow!25} 0 & \cellcolor{yellow!25} 0 & \cellcolor{red!25} 29 & \cellcolor{green!25} 225 & \cellcolor{yellow!25} 248 & \cellcolor{red!20} 3 & \cellcolor{red!20} 16 & \cellcolor{green!25} 394 & \cellcolor{red!25} 11 & \cellcolor{green!25} 478 & \cellcolor{green!25} 127 & \cellcolor{red!20} 8 \\
             \bottomrule
         \end{tabular}
     }
     \vspace{-0.2cm}
 \end{table*}

 \begin{table*}[ht!]
     \centering
     \caption{
Quality Metrics of Different Models Benchmarked on \multipleJava (\ie HumanEval) and \mcJava. Cyc refers to Cyclomatic Complexity, while CoC denotes Cognitive Complexity}     \normalsize
     \label{tab:model-performance-java}
     \resizebox{\linewidth}{!}{
         \begin{tabular}{llcc|cccccc|cccccc}
             \toprule
             \textbf{Dataset} & \textbf{Model} & \textbf{Precision} & \textbf{Pass@1} & \multicolumn{6}{c|}{\textbf{SonarCloud Metrics}} & \multicolumn{6}{c}{\textbf{PMD Metrics}} \\
             \cmidrule(lr){5-10} \cmidrule(lr){11-16}
             & & & & \textbf{LoC} & \textbf{Security Hotspot} & \textbf{Reliability} & \textbf{Maintainability} & \textbf{CyC} & \textbf{CoC} & \textbf{Best Practices} & \textbf{Code Style} & \textbf{Design} & \textbf{Error Prone} & \textbf{Multithreading} & \textbf{Performance} \\
             \midrule
             
             \multirow{12}{*}{\centering \emph{MultipLE}-Java} & \multirow{2}{*}{CodeLllama 7B} & \cellcolor[gray]{.75} 16 bit & \cellcolor[gray]{.75} 0.25 & \cellcolor[gray]{.75} 4640 & \cellcolor[gray]{.75} 0 & \cellcolor[gray]{.75} 24 & \cellcolor[gray]{.75} 946 & \cellcolor[gray]{.75} 824 & \cellcolor[gray]{.75} 687 & \cellcolor[gray]{.75} 424 & \cellcolor[gray]{.75} 9245 & \cellcolor[gray]{.75} 445 & \cellcolor[gray]{.75} 53 & \cellcolor[gray]{.75} 0 & \cellcolor[gray]{.75} 20 \\
            
             & & \cellcolor[gray]{.95} 4 bit & \cellcolor[gray]{.95} 0.24 & \cellcolor[gray]{.95} 4434 $\darkred\downarrow$ & \cellcolor[gray]{.95} 0 & \cellcolor{yellow!25} 24 & \cellcolor{green!25} 804 & \cellcolor{green!25} 714 & \cellcolor{green!25} 539 & \cellcolor{green!25} 417 & \cellcolor{green!25} 8952 & \cellcolor{green!25} 440 & \cellcolor{red!25} 56 & \cellcolor{yellow!25} 0 & \cellcolor{red!25} 25 \\
             
             & \multirow{2}{*}{CodeLllama 13B} & \cellcolor[gray]{.75} 16 bit & \cellcolor[gray]{.75} 0.29 & \cellcolor[gray]{.75} 4720  & \cellcolor[gray]{.75} 0 & \cellcolor[gray]{.75} 26 & \cellcolor[gray]{.75} 937 & \cellcolor[gray]{.75} 809 & \cellcolor[gray]{.75} 706 & \cellcolor[gray]{.75} 441 & \cellcolor[gray]{.75} 9620 & \cellcolor[gray]{.75} 464 & \cellcolor[gray]{.75} 89 & \cellcolor[gray]{.75} 0 & \cellcolor[gray]{.75} 25 \\
            
             & & \cellcolor[gray]{.95} 4 bit & \cellcolor[gray]{.95} 0.29 & \cellcolor[gray]{.95} 4593 $\darkred\downarrow$ & \cellcolor[gray]{.95} 0 & \cellcolor{green!25} 24 & \cellcolor{green!25} 862 & \cellcolor{green!25} 778 & \cellcolor{green!25} 630 & \cellcolor{red!25} 444 & \cellcolor{green!25} 9212 & \cellcolor{green!25} 439 & \cellcolor{green!25} 83 & \cellcolor{yellow!25} 0 & \cellcolor{green!25} 24 \\
             
             & \multirow{2}{*}{CodeLllama 34B} & \cellcolor[gray]{.75} 16 bit & \cellcolor[gray]{.75} 0.35 & \cellcolor[gray]{.75} 4478 & \cellcolor[gray]{.75} 0 & \cellcolor[gray]{.75} 24 & \cellcolor[gray]{.75} 845 & \cellcolor[gray]{.75} 743 & \cellcolor[gray]{.75} 690 & \cellcolor[gray]{.75} 419 & \cellcolor[gray]{.75} 9034 & \cellcolor[gray]{.75} 441 & \cellcolor[gray]{.75} 75 & \cellcolor[gray]{.75} 0 & \cellcolor[gray]{.75} 14 \\
            
             & & \cellcolor[gray]{.95} 4 bit & \cellcolor[gray]{.95} 0.34 & \cellcolor[gray]{.95} 4704 $\darkgreen\uparrow$ & \cellcolor[gray]{.95} 0 & \cellcolor{red!25} 25 & \cellcolor{red!25} 913 & \cellcolor{red!25} 804 & \cellcolor{green!25} 660 & \cellcolor{red!25} 426 & \cellcolor{red!25} 9271 & \cellcolor{red!25} 442 & \cellcolor{red!30} 102 & \cellcolor{red!25} 1 & \cellcolor{red!25} 31 \\
             \cline{2-16}
             
             & \multirow{2}{*}{DeepSeek-Coder 1.3B} & \cellcolor[gray]{.75} 16 bit & \cellcolor[gray]{.75} 0.40 & \cellcolor[gray]{.75} 4441 & \cellcolor[gray]{.75} 0 & \cellcolor[gray]{.75} 31 & \cellcolor[gray]{.75} 827 & \cellcolor[gray]{.75} 722 & \cellcolor[gray]{.75} 541 & \cellcolor[gray]{.75} 396 & \cellcolor[gray]{.75} 9102 & \cellcolor[gray]{.75} 431 & \cellcolor[gray]{.75} 87 & \cellcolor[gray]{.75} 0 & \cellcolor[gray]{.75} 11 \\
            
             & & \cellcolor[gray]{.95} 4 bit & \cellcolor[gray]{.95} 0.37 & \cellcolor[gray]{.95} 4406 \textbf{$\darkred\downarrow$} & \cellcolor[gray]{.95} 0 & \cellcolor{green!25} 24 & \cellcolor{red!25} 960 & \cellcolor{red!25} 728 & \cellcolor{green!25} 539 & \cellcolor{red!25} 420 & \cellcolor{red!25} 9380 & \cellcolor{green!25} 425 & \cellcolor{green!25} 46 & \cellcolor{red!25} 1 & \cellcolor{green!25} 4 \\
             
             & \multirow{2}{*}{DeepSeek-Coder 6.7B} & \cellcolor[gray]{.75} 16 bit & \cellcolor[gray]{.75} 0.56 & \cellcolor[gray]{.75} 4579 & \cellcolor[gray]{.75} 0 & \cellcolor[gray]{.75} 24 & \cellcolor[gray]{.75} 904 & \cellcolor[gray]{.75} 719 & \cellcolor[gray]{.75} 590 & \cellcolor[gray]{.75} 426 & \cellcolor[gray]{.75} 9546 & \cellcolor[gray]{.75} 431 & \cellcolor[gray]{.75} 78 & \cellcolor[gray]{.75} 2 & \cellcolor[gray]{.75} 11 \\
            
             & & \cellcolor[gray]{.95} 4 bit & \cellcolor[gray]{.95} 0.55 & \cellcolor[gray]{.95} 4523 $\darkred\downarrow$ & \cellcolor[gray]{.95} 0 & \cellcolor{red!25} 31 & \cellcolor{red!25} 964 & \cellcolor{red!25} 776 & \cellcolor{green!25} 533 & \cellcolor{green!25} 424 & \cellcolor{green!25} 9417 & \cellcolor{green!25} 424 & \cellcolor{green!25} 68 & \cellcolor{red!25} 3 & \cellcolor{green!25} 8 \\
             
             & \multirow{2}{*}{DeepSeek-Coder 33B} & \cellcolor[gray]{.75} 16 bit & \cellcolor[gray]{.75} 0.54 & \cellcolor[gray]{.75} 4556 & \cellcolor[gray]{.75} 0 & \cellcolor[gray]{.75} 24 & \cellcolor[gray]{.75} 853 & \cellcolor[gray]{.75} 738 & \cellcolor[gray]{.75} 569 & \cellcolor[gray]{.75} 419 & \cellcolor[gray]{.75} 9539 & \cellcolor[gray]{.75} 433 & \cellcolor[gray]{.75} 75 & \cellcolor[gray]{.75} 2 & \cellcolor[gray]{.75} 13 \\
            
             & & \cellcolor[gray]{.95} 4 bit & \cellcolor[gray]{.95} 0.54 & \cellcolor[gray]{.95} 4513 $\darkred\downarrow$ & \cellcolor[gray]{.95} 0 & \cellcolor{yellow!25} 24 & \cellcolor{green!25} 851 & \cellcolor{green!25} 736 & \cellcolor{green!25} 550 & \cellcolor{green!25} 418 & \cellcolor{green!25} 9518 & \cellcolor{red!25} 438 & \cellcolor{red!25} 76 & \cellcolor{red!25} 3 & \cellcolor{red!25} 30 \\
             \midrule
            
             \multirow{12}{*}{\centering McEval-Java} & \multirow{2}{*}{CodeLllama 7B} & \cellcolor[gray]{.75} 16 bit & \cellcolor[gray]{.75} 0.28 & \cellcolor[gray]{.75} 1322 & \cellcolor[gray]{.75} 0 & \cellcolor[gray]{.75} 9 & \cellcolor[gray]{.75} 326 & \cellcolor[gray]{.75} 252 & \cellcolor[gray]{.75} 248 & \cellcolor[gray]{.75} 115 & \cellcolor[gray]{.75} 349 & \cellcolor[gray]{.75} 55 & \cellcolor[gray]{.75} 14 & \cellcolor[gray]{.75} 2 & \cellcolor[gray]{.75} 5 \\
            
             & & \cellcolor[gray]{.95} 4 bit & \cellcolor[gray]{.95} 0.26 & \cellcolor[gray]{.95} 1457 $\darkgreen\uparrow$ & \cellcolor[gray]{.95} 0 & \cellcolor{yellow!25} 9 & \cellcolor{red!20}  364 & \cellcolor{red!20} 303 & \cellcolor{red!25} 308 & \cellcolor{red!25} 117 & \cellcolor{red!20} 422 & \cellcolor{red!25} 59 & \cellcolor{green!25} 8 & \cellcolor{red!25} 4 & \cellcolor{green!25} 3 \\
            
             & \multirow{2}{*}{CodeLllama 13B} & \cellcolor[gray]{.75} 16 bit & \cellcolor[gray]{.75} 0.26 & \cellcolor[gray]{.75} 1559 & \cellcolor[gray]{.75} 0 & \cellcolor[gray]{.75} 9 & \cellcolor[gray]{.75} 382 & \cellcolor[gray]{.75} 308 & \cellcolor[gray]{.75} 326 & \cellcolor[gray]{.75} 136 & \cellcolor[gray]{.75} 441 & \cellcolor[gray]{.75} 59 & \cellcolor[gray]{.75} 13 & \cellcolor[gray]{.75} 2 & \cellcolor[gray]{.75} 6 \\
            
             & & \cellcolor[gray]{.95} 4 bit & \cellcolor[gray]{.95} 0.26 & \cellcolor[gray]{.95} 1378 $\darkred\downarrow$ & \cellcolor[gray]{.95} 0 & \cellcolor{yellow!25} 9 & \cellcolor{green!20} 348 & \cellcolor{green!20} 252 & \cellcolor{green!20} 246 & \cellcolor{green!25} 119 & \cellcolor{green!20} 367 & \cellcolor{green!25} 54 & \cellcolor{green!25} 9 & \cellcolor{green!25} 1 & \cellcolor{green!25} 3 \\
            
             & \multirow{2}{*}{CodeLllama 34B} & \cellcolor[gray]{.75} 16 bit & \cellcolor[gray]{.75} 0.39 & \cellcolor[gray]{.75} 1351 & \cellcolor[gray]{.75} 0 & \cellcolor[gray]{.75} 9 & \cellcolor[gray]{.75} 356 & \cellcolor[gray]{.75} 248 & \cellcolor[gray]{.75} 227 & \cellcolor[gray]{.75} 121 & \cellcolor[gray]{.75} 368 & \cellcolor[gray]{.75} 51 & \cellcolor[gray]{.75} 15 & \cellcolor[gray]{.75} 3 & \cellcolor[gray]{.75} 7 \\
            
             & & \cellcolor[gray]{.95} 4 bit & \cellcolor[gray]{.95} 0.37 & \cellcolor[gray]{.95} 1316 $\darkred\downarrow$ & \cellcolor[gray]{.95} 0 & \cellcolor{yellow!25} 9 & \cellcolor{green!25} 335 & \cellcolor{green!25} 238 & \cellcolor{green!25} 204 & \cellcolor{green!25} 115 & \cellcolor{green!25} 363 & \cellcolor{red!25} 53 & \cellcolor{red!25} 17 & \cellcolor{yellow!25} 3 & \cellcolor{green!20} 2 \\
             \cline{2-16}
             
             & \multirow{2}{*}{DeepSeek-Coder 1.3B} & \cellcolor[gray]{.75} 16 bit & \cellcolor[gray]{.75} 0.37 & \cellcolor[gray]{.75} 1444 & \cellcolor[gray]{.75} 0 & \cellcolor[gray]{.75} 9 & \cellcolor[gray]{.75} 339 & \cellcolor[gray]{.75} 277 & \cellcolor[gray]{.75} 272 & \cellcolor[gray]{.75} 122 & \cellcolor[gray]{.75} 375 & \cellcolor[gray]{.75} 52 & \cellcolor[gray]{.75} 10 & \cellcolor[gray]{.75} 5 & \cellcolor[gray]{.75} 5 \\
            
             & & \cellcolor[gray]{.95} 4 bit & \cellcolor[gray]{.95} 0.35 & \cellcolor[gray]{.95} 1472 $\darkgreen\uparrow$ & \cellcolor[gray]{.95} 0 & \cellcolor{yellow!25} 9 & \cellcolor{red!25} 346 & \cellcolor{yellow!25} 277 & \cellcolor{red!25} 299 & \cellcolor{green!25} 119 & \cellcolor{red!25} 388 & \cellcolor{red!25} 60 & \cellcolor{red!20} 23 & \cellcolor{yellow!25} 5 & \cellcolor{red!25} 7 \\
             
             & \multirow{2}{*}{DeepSeek-Coder 6.7B} & \cellcolor[gray]{.75} 16 bit & \cellcolor[gray]{.75} 0.47 & \cellcolor[gray]{.75} 1436 & \cellcolor[gray]{.75} 0 & \cellcolor[gray]{.75} 9 & \cellcolor[gray]{.75} 357 & \cellcolor[gray]{.75} 276 & \cellcolor[gray]{.75} 289 & \cellcolor[gray]{.75} 114 & \cellcolor[gray]{.75} 384 & \cellcolor[gray]{.75} 53 & \cellcolor[gray]{.75} 7 & \cellcolor[gray]{.75} 4 & \cellcolor[gray]{.75} 8 \\
            
             & & \cellcolor[gray]{.95} 4 bit & \cellcolor[gray]{.95} 0.60 & \cellcolor[gray]{.95} 1376 $\darkred\downarrow$ & \cellcolor[gray]{.95} 0 & \cellcolor{yellow!25} 9 & \cellcolor{green!25} 337 & \cellcolor{green!25} 257 & \cellcolor{green!25} 232 & \cellcolor{red!25} 116 & \cellcolor{green!25} 373 & \cellcolor{green!25} 51 & \cellcolor{red!25} 13 & \cellcolor{green!25} 2 & \cellcolor{green!20} 3 \\
             
             & \multirow{2}{*}{DeepSeek-Coder 33B} & \cellcolor[gray]{.75} 16 bit & \cellcolor[gray]{.75} 0.49 & \cellcolor[gray]{.75} 1339 & \cellcolor[gray]{.75} 0 & \cellcolor[gray]{.75} 9 & \cellcolor[gray]{.75} 347 & \cellcolor[gray]{.75} 245 & \cellcolor[gray]{.75} 252 & \cellcolor[gray]{.75} 115 & \cellcolor[gray]{.75} 357 & \cellcolor[gray]{.75} 56 & \cellcolor[gray]{.75} 7 & \cellcolor[gray]{.75} 3 & \cellcolor[gray]{.75} 5 \\
            
             & & \cellcolor[gray]{.95} 4 bit & \cellcolor[gray]{.95} 0.62 & \cellcolor[gray]{.95} 1465 $\darkgreen\uparrow$ & \cellcolor[gray]{.95} 1 & \cellcolor{yellow!25} 9 & \cellcolor{green!25} 343 & \cellcolor{red!25} 280 & \cellcolor{red!25} 274 & \cellcolor{red!25} 118 & \cellcolor{red!25} 387 & \cellcolor{green!25} 55 & \cellcolor{red!25} 13 & \cellcolor{green!25} 2 & \cellcolor{red!25} 6 \\
             \bottomrule
         \end{tabular}
     }
     \vspace{-0.2cm}
 \end{table*}

To ensure a clear and structured discussion, we begin by presenting the results of our quantitative analysis. We then follow with qualitative examples that highlight notable behaviors of both quantized and non-quantized large code models, offering insights that inform future research and might reveal emerging trends.

\tabref{tab:model-performance-py} presents a comprehensive overview of SonarCloud quality metrics, Pylint, and Flake8 for 12 model configurations, comparing full-precision (FP) and quantized models across different sizes. The upper section of the table summarizes results from evaluations on \multipleE (\ie HumanEval for Python), while the lower section reports findings from benchmarking the models on the \mcEval dataset, which also focuses on Python. The results are organized by programming language to ensure a consistent and meaningful comparison, as the static analysis tools employed in the study are specifically designed for either Java or Python, each with distinct evaluation criteria and quality assessment methodologies. Furthermore, to facilitate comparisons, we use three different background colors for the cells: 
\begin{itemize}

 	\item A light green color \raisebox{0.1cm}{\tikz[baseline=(char.base)]{\node[fill=green!25, draw=green!25, minimum width=0.2cm, minimum height=0.3cm, inner sep=6pt] (char) {};}} represents a positive improvement in the given metric for the quantized model. For instance, CodeLlama 7B exhibits fewer code smells compared to its non-quantized counterpart—hence, the value 28 is highlighted.
	\item A light red color \raisebox{0.1cm}{\tikz[baseline=(char.base)]{\node[fill=red!25, draw=red!25, minimum width=0.2cm, minimum height=0.3cm, inner sep=6pt] (char) {};}} signifies a negative impact on the quantized model for that metric. For example, the quantized version of CodeLlama 13B introduces more bugs than its full-precision equivalent.
	\item A light yellow color \raisebox{0.1cm}{\tikz[baseline=(char.base)]{\node[fill=yellow, draw=yellow, minimum width=0.2cm, minimum height=0.3cm, inner sep=6pt] (char) {};}} denotes a neutral effect, indicating no significant variation in the metric.
\end{itemize}

To begin with, from \tabref{tab:model-performance-py}, we derive the total number of issues for each model by summing \emph{Security-Hotspots, Reliability issues (\ie bug incidence), Maintainability Issues (\ie Code Smells), CyC, CoC} for both the CodeLlama and DeepSeek-Coder families. 

We refrain from including the plots due to space limitations. Nonetheless, these can be found in our replication package \cite{replication}.

Before providing specific insights, from a general standpoint,  we find that the behavior of CodeLlama and DeepSeek-Coder across different configurations, languages, and benchmarks does not exhibit significant variation. To this extent, if we closely inspect the benchmark \mcEval, we observe that, on average, quantized models produce code with 2,892 quality issues, as assessed by SonarCloud. In contrast, non-quantized models generate Python code with a total of 3,084 quality issues. This suggests that, at least for \mcPython, 4-bit AWQ quantized models withstand the impact of compression without a significant decline in code quality--quite the contrary, with a notable reduction that accounts for ~6\% in favor of quantized models.

For Java, the trend shifts slightly, with quantized models generating 5,347 quality issues compared to 5,381 in non-quantized models. This represents a marginal reduction, suggesting that quantization has a minimal yet consistent effect on maintaining code quality across different programming languages.


While this result may seem surprising at first, we remind the reader that quantization primarily affects model precision rather than its overall ability to generate syntactically and semantically valid code. Additionally, the resilience observed in quantized models may stem from the robustness of AWQ, which selectively preserves key activations that contribute to meaningful code generation.

This observed behavior of quantized models prompts a deeper examination of their impact on specific code quality metrics and static features. Turning to CodeLlama (upper section of \tabref{tab:model-performance-py}), we observe a clear downward trend in total quality issues as model size increases. Notably, the 34B variant exhibits the fewest quality issues among CodeLlama models, regardless of whether it operates in full-precision or quantized mode.
However, the effect of quantization varies significantly across model sizes, influencing the quality of automatically generated code in different ways. For instance, the quantized CodeLlama 7B model produces higher-quality code than its full-precision counterpart, exhibiting fewer code smells (as reflected in improved maintainability scores) and a lower likelihood of generating buggy code. This suggests that, despite operating with compressed knowledge, the quantized version of CodeLlama 7B remains robust, retaining its ability to generate reliable and maintainable code.

As model size increases, quantization appears to affect the ability of the model to generate code with quality comparable to its full-precision counterpart. Notably, quantized models tend to generate slightly longer code, with the most pronounced increase of 50 lines observed in the 13B variant. Security-Hotspots remain largely unaffected by quantization, while the impact on reliability—measured through the number of detected bugs—does not follow a clear pattern across different model configurations.

Finally, in terms of maintainability, assessed through the number of code smells, quantization slightly reduces maintainability scores in smaller models (7B and 13B), whereas its effect on the larger 34B model is negligible. This suggests that, while quantization can have a mild influence on maintainability in smaller models, its overall impact diminishes as model size increases.


When examining Pylint and Flake8 metrics, we find that the trends observed in SonarCloud largely hold true, with no major deviations. In a similar fashion to the metrics computed using SonarCloud, also for Pylint and Flake8, there appears to be an inverse correlation of metrics with the model's size. In other words, as the number of parameters increases ($\darkgreen\Uparrow$), the number of quality issues flagged by state-of-the-art static analysis tools decreases ($\darkred\Downarrow$). 

A comparative analysis between models reveals a distinct performance gap between the CodeLlama family and DeepSeek-Coder when also accounting for the ability of the model to produce correct code (\ie Pass@1) across different model sizes. Regardless of parameter count, DeepSeek-Coder consistently produces more reliable code with fewer flagged issues than CodeLlama, even as the latter scales up. This demonstrates that DeepSeek-Coder not only generates higher-quality code but also achieves a higher pass@1 rate, showcasing better coding skills in producing functionally correct and maintainable code across its various configurations.

The only notable exception is the behavior of the DeepSeek-Coder 33B quantized model, which unexpectedly exhibits an increase in maintainability issues compared to its full-precision counterpart. This holds across both benchmarks.

As for Java, the complete set of metrics is reported in \tabref{tab:model-performance-java}. In particular, we include the same SonarCloud metrics as used for Python, with the addition of PMD--as mentioned in \secref{sec:implementation}--to help broaden the scope of our evaluation and enhance the generalizability of our findings.

Examining the SonarCloud metrics, we observe a particular trend in CodeLlama’s quantized models, which generally produce longer, more complex, and less readable code. This pattern is particularly pronounced when CodeLlama is tasked with generating Java code from the \multipleJava benchmark. A similar trend emerges in the DeepSeek family of models, especially when scaling up in size. Specifically, when evaluating the \mcJava benchmark, DeepSeek-Coder exhibits characteristics akin to those observed in CodeLlama on \multipleJava. 

While these observations alone may not fully explain the effects of quantization, they gain deeper significance when cross-referenced with functional correctness. Across various parameter configurations, DeepSeek-Coder consistently produces code that successfully passes test cases, reinforcing the pattern previously observed in Python. 


Examining the PMD metrics, we identify a consistent trend across both quantized and non-quantized models: larger models tend to produce code with more design issues. This is likely due to their greater capacity for generating longer code, which in turn increases cyclomatic complexity. The relationship between model size, lines of code, and structural complexity reinforces the notion that while larger models exhibit greater expressive power, they may also introduce inefficiencies in code organization and maintainability.

Building on this observation, the evaluation of performance-related attributes in generated code reveals a notable divergence between quantized and non-quantized models. In general, performance issues tend to escalate as model size increases. However, while this pattern is not always evident in non-quantized models, it becomes more pronounced in their quantized counterparts. Remarkably, CodeLlama 34B and DeepSeek-Coder 33B on \multipleJava exhibit more than twice the number of performance-related issues compared to their full-precision versions. In contrast, our analysis of \mcEval\ reveals no consistent pattern, making it difficult to even speculate about the effect of quantization on performance within this benchmark.

Ultimately, the key factor that enables us to draw definitive conclusions with a good degree of confidence is the outcome of the statistical tests conducted. Across both benchmarks and programming languages, we identify \texttt{Warning} from \textit{Pylint} and \texttt{Code Style} from \textit{PMD} metrics, where significant differences emerge, yet with an effect size ranging from negligible (N) to (S). For all other comparisons, particularly those evaluated using the Wilcoxon test, the observed differences are not statistically significant.

The results of the statistical analysis are included in the replication package \cite{replication}.

Finally, the manual analysis results indicate a Kappa Score of 0.8416 for the consistency metric, demonstrating a strong level of agreement between annotators. In contrast, the readability metric yielded a lower Kappa Score of 0.69, reflecting moderate to high inter-rater agreement. This discrepancy suggests that while consistency assessments remained relatively objective, readability evaluations introduced a higher degree of subjectivity. A closer look at the number of conflicts further supports this observation: 7 cases (14\%) of disagreement arose for consistency, whereas 14 cases (28\%) were recorded for readability—double the amount. These conflicts were resolved through open discussions between the annotators, ensuring a consensus-driven refinement of the final labels. Following the resolution of conflicts, the final distribution of classifications for consistency was as follows: 55 labeled as acceptable, 27 as good, and 18 as poor. For readability, the distribution was 45 acceptable, 28 good, and 27 poor.
These findings suggest that quantization does not significantly disrupt the model’s ability to generate coherent code—as consistency remains largely preserved. However, it does seem to introduce subtle degradations in readability, likely due to less optimal identifier choices, reduced clarity in variable naming, and structurally more convoluted code patterns.


With that in mind, we summarize our key insights as follows:

\begin{tcolorbox}[colback=gray!10, colframe=black, arc=5mm, title={Summary of Experimental Outcomes:}]

\rev{Our initial hypothesis posited that quantization, while preserving functional correctness, would degrade certain software quality attributes (\eg maintainability and reliability). This expectation motivated the central research question of our study, which seeks to provide a comprehensive examination of code quality under the influence of quantization. In addition, we extend our analysis across models of varying sizes to investigate how model scale affects this relationship and to explore whether smaller, quantized models exhibit a more pronounced quality gap compared to their larger counterparts.}
Our findings indicate that larger models generally exhibit fewer quality issues, irrespective of whether they are quantized. Specifically, security remains stable, while reliability and maintainability show only minor fluctuations. However, quantization does have a modest impact on readability and maintainability, primarily due to a slight increase in the length of the generated code. Overall, quantization can be applied to LCMs without significant degradation in quality, though practical trade-offs should be carefully weighed.

\end{tcolorbox}

\subsection{Qualitative Examples}

\begin{figure}[ht!]
	\centering
	\includegraphics[width=1\columnwidth]{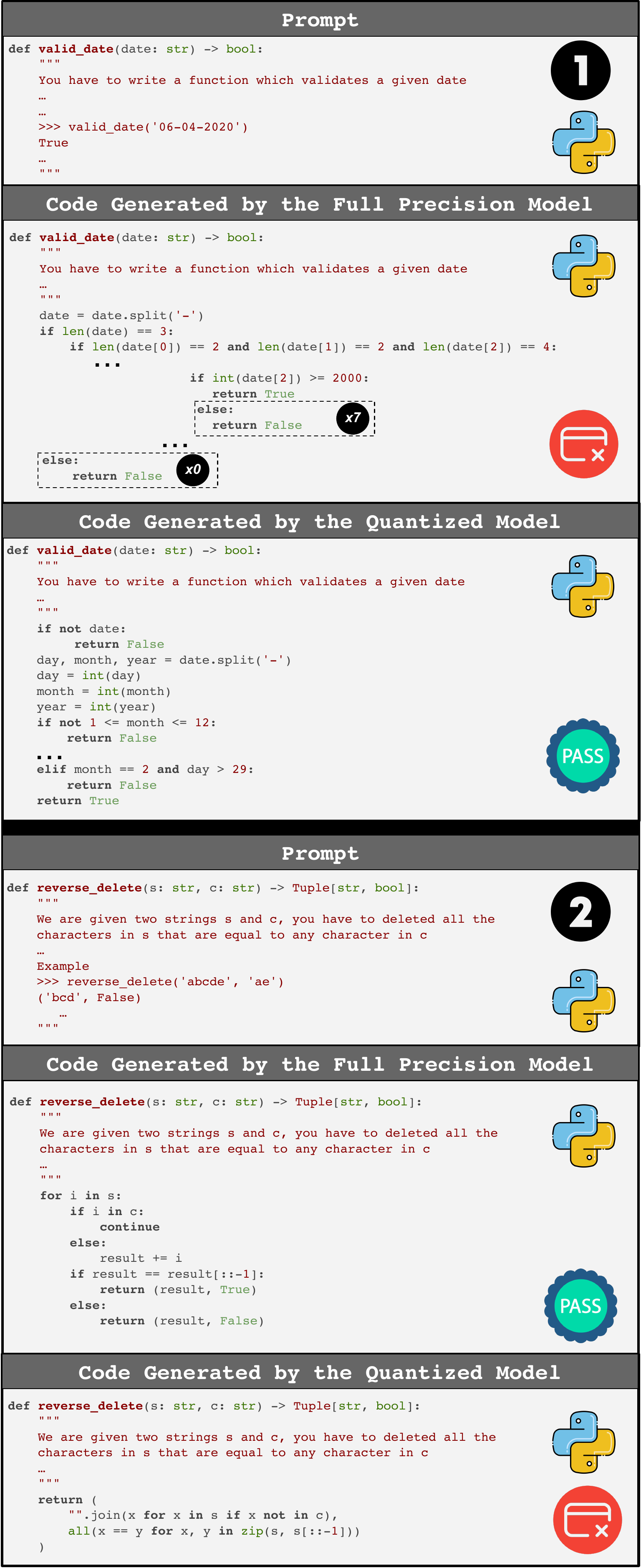}
	\caption{Example of four distinct implementations of two different given requirements in Python.}
	\label{fig:examples}
\end{figure}

We conclude our discussion by examining two contrasting cases from DeepSeek-Coder 33B predictions in 16-bit (FP) and 4-bit (quantized, AWQ) settings on \multiplePython (\figref{fig:examples}), offering valuable insights that can guide future research on the topic.

In \circled{1}, the task is to validate a date. The FP model generates a verbose, incorrect, and complex solution with a higher Cyclomatic Complexity (CyC: 10), making the code less maintainable and more error-prone. Conversely, the quantized model produces a concise, correct, and structured implementation with lower complexity (CyC: 8), demonstrating that quantization can enhance efficiency by discouraging unnecessary complexity. However, in \circled{2}, where the task is to remove characters in s that appear in c, the trend reverses. The FP model generates a clear and well-structured solution, whereas the quantized model introduces unnecessary complexity through a nested loop and an inefficient palindrome check, leading to an O(n) time and space overhead. Moreover, the palindrome check is applied incorrectly to s instead of the filtered result, reducing both correctness and readability.

These cases highlight the dual impact of quantization on code generation. In some scenarios, such as \circled{1}, quantization improves efficiency and maintainability by eliminating redundant complexity. However, in others, like \circled{2}, it can introduce inefficiencies and errors, ultimately degrading code quality. This trade-off raises concerns regarding quantization’s role in the sustainability of large code models. While reducing model size improves efficiency, it is crucial to ensure that this process does not lead to more harm than good by compromising code quality. Further exploration of trade-offs is necessary to prevent unintended performance bottlenecks before quantization can be established as a sustainable and effective technique for large code models. A deeper understanding of these trade-offs will be crucial in optimizing quantization strategies while preserving model efficiency and software quality attributes.


\section{Threats to Validity}
\label{sec:threats}

\textbf{Construct validity} threats pertain to the accuracy of measuring theoretical concepts through empirical methods, particularly the metrics used to address our research questions. In our study, a key risk involves the choice of static analysis tools for assessing the quality of generated code. To mitigate this, as detailed in \secref{sec:design}, we rely on well-established static analysis tools and the standardized metrics they compute to evaluate automatically generated code. Additionally, we evaluate code quality across two different programming languages and multiple benchmarks, which helps reduce potential biases tied to language-specific or domain-specific characteristics. Yet, we acknowledge that outcomes may differ when applied to other programming languages.

\textbf{Internal validity} threats pertain to factors within the study that may influence the results. One potential concern is the selection of models used for analysis. To address this, we chose state-of-the-art models that have been extensively used in prior research \cite{virk2024enhancing,zhu2024effectiveness,yang2024synthesizing,li2023structured}, ensuring relevance and robustness in our findings. 

Another risk involves the choice of quantization methods. To this end, we carefully reviewed existing literature and adhered to best practices in quantization research, following established methodologies to ensure comparability and validity of results.

Finally, to account for output variability, we generated ten predictions per instance. For this analysis, we focused on the smallest models from the CodeLlama (7B) and DeepSeek-Coder (1.3B) families on the McEval benchmark. We limited our scope to these smaller models due to the substantial computational cost of repeated sampling with larger models. Nonetheless, this conservative choice provides a lower-bound estimate of variability. Using Friedman’s test to isolate iteration-level from task-level effects, we found no statistically significant differences in the evaluated quality metrics.

\textbf{Conclusion validity} threats relate to the connection between our experimental process and the reliability of our conclusions. We mitigate these threats by employing appropriate statistical methods where necessary, ensuring that our findings are statistically sound and reproducible.

\textbf{External validity} threats concern the generalizability of our findings beyond the scope of our study. While our analysis focuses on code generation as a representative software engineering task, results may vary for other code-related applications.

\section{Conclusions and Future Work} 
\label{sec:conclusions}

In this study, we empirically investigated the impact of quantizing large code models beyond functional correctness, emphasizing its effects on qualitative aspects and static properties of generated code. To ensure a comprehensive assessment, we leveraged a suite of state-of-the-art static analysis tools, including SonarCloud, PMD, Pylint, and Flake8, to capture a broad spectrum of code quality metrics.

Our findings reveal that quantization \textbf{is not a deal-breaker}—its ability to produce functionally correct code extends to qualitative aspects and static properties as well. This result largely contradicts our initial hypothesis, which assumed that the loss of information induced by quantization would degrade software quality. 

However, while quantization does not entirely compromise the quality of generated code, its impact can be nuanced, as illustrated in \figref{fig:examples}. In certain cases, it enhances efficiency and maintainability by minimizing unnecessary complexity, whereas in others, it may introduce inefficiencies or errors. This indicates that adopting model compression techniques for large code models requires careful consideration of their trade-offs. Our future research agenda aims to explore the long-term implications of using quantized models in software development activities to promote sustainability. With a broader understanding of quantization from a holistic perspective, we can delve into its tangible benefits in real-world scenarios and assess the trade-offs that developers and practitioners are willing to accept. 


\section{Acknowledgments}
This research was partially funded by NSF CCF-2451058. A complete list of image attributions can be found at \cite{replication}.

\balance
\bibliographystyle{IEEEtran}
\bibliography{main}

\end{document}